# The Solutions to Uncertainty Problem of Urban Fractal Dimension Calculation

Yanguang Chen

(Department of Geography, College of Urban and Environmental Sciences, Peking University, Beijing 100871, P.R. China. E-mail: chenyg@pku.edu.cn)

**Abstract**: Fractal geometry provides a powerful tool for scale-free spatial analysis of cities, but the fractal dimension calculation results always depend on methods and scopes of study area. This phenomenon has been puzzling many researchers. This paper is devoted to discussing the problem of uncertainty of fractal dimension estimation and the potential solutions to it. Using regular fractals as archetypes, we can reveal the causes and effects of the diversity of fractal dimension estimation results by analogy. The main factors influencing fractal dimension values of cities include prefractal structure, multi-scaling fractal patterns, and self-affine fractal growth. The solution to the problem is to substitute the real fractal dimension values with comparable fractal dimensions. The main measures are as follows: First, select a proper method for a special fractal study. Second, define a proper study area for a city according to a study aim, or define comparable study areas for different cities. These suggestions may be helpful for the students who takes interest in or even have already participated in the studies of fractal cities.

**Key words**: Fractal; prefractal; multifractals; self-affine fractals; fractal cities; fractal dimension measurement

# 1. Introduction

A scientific research is involved with two processes: description and understanding. A study often proceeds first by describing how a system works and then by understanding why (Gordon, 2005). Scientific description relies heavily on measurement and mathematical modeling, and scientific explanation is mainly to use observation, experience, and experiment (Henry, 2002). Mathematical reasoning and systematic controlled experiment represent two bases of great achievements in the development of Western science (Einstein, 1962). The basic method of description is measurement, which form a link between mathematical modeling and empirical studies (Taylor, 1983). The precondition of effective description by measurement and mathematical modeling is to find a characteristic scale, which always takes on a 1-dimension measure and termed *characteristic length* (Mandelbrot, 1982; Hao, 1986; Liu and Liu, 1993; Takayasu, 1990; Wang and Li, 1996). However, for complex systems such as a city or system of cities, it is difficult or even impossible to find a characteristic scale to make mathematical model or quantitative analysis. In this case, it is an advisable selection to replace characteristic scale with scaling notion.

Fractal geometry provides a powerful tool of scaling analysis for complex systems, and it is useful in both theoretical and empirical research on cities. A number of studies on fractal cities verified the effect and power of fractal methods and fractal parameters (see e.g. Ariza-Villaverde *et al*, 2013; Batty and Longley, 1994; Benguigui *et al*, 2010; Chen, 2008; Chen and Wang, 2013; Feng and Chen, 2010; Encarnação *et al*, 2013; Frankhauser, 1994; Gao *et al*, 2017; Hu *et al*, 2012; Jiang and Brandt, 2016; Longley *et al*, 1991; Murcio *et al*, 2015; Tannier *et al*, 2011;Thomas *et al*, 2007; Thomas *et al*, 2010). Unfortunately, new problems have arisen in recent years, that is, the results of fractal dimension estimation for urban form depend on scope of study area and methods of fractal dimension measurement and calculation (see e.g. Batty and Longley, 1994; Benguigui *et al*, 2000; Chen, 2008; Feng and Chen, 2010; Frankhauser, 2014). This puzzles many scholars who take interest in fractal cities for a long time. What is the root of the kind of problems? Opinions differ in the academic circle, and different scholars give different viewpoints. This paper is devoted to revealing the possible reasons for the diversity of fractal dimension calculation results in the fractal studies on cities. As a preparation, five concepts are explained in advance as below: (1) *scaling range*: the middle straight line segments in log-log plots reflect the relationships between the linear scales of measurement and the corresponding measures, and the slopes of the segments indicate fractal dimension; (2) *real fractals*: scaling range is infinite, and Lebesgue measure is zero; (3) *prefractals*: real fractals: scaling range is finite, and Lebesgue measure is greater than zero; (4) *monofractals*, or *unifractals*: there is one scaling process; (5) *multifractals*: there are more than one scaling process. The rest parts are organized as follows. In Section 2, the relationships between measurements and fractal dimension are explained. In Section 3, various methods of fractal dimension estimation are collected and sorted. In Section 4, several related questions are discussed. Finally, the conclusions are drawn by summarizing the main points of this work.

## 2. Measurement and dimension

### 2.1 Euclidean measurement and fractal dimension

As indicated above, scientific research begins with a description. In order to describe a thing, we should measure its number (for a point set), or length (for a line), or area (for a surface), or volume (for a body). Using number, length, area, or volume, we can define a measurement such as density and shape index (Haggett *et al*, 1977). In this way, the characters of a system can be condensed into a simple number. As Lord Kelvin once pointed out: "*When you can measure what you are speaking about, and express it in numbers, you know something about it; but when you cannot measure it, when you cannot express it in numbers, your knowledge is of a meager and unsatisfactory kind.*" (Cited from Taylor, 1983, p37) Therefore, Edwards Deming said, "*In God we trust, all others bring data*" (Cited from Hastie *et al*, 2016). Once a number is obtained for a thing by measurement, some uncertainty is eliminated. Then we will say we gain information from it. Anyway, information indicates an increase of understanding and resolution of uncertainty (Shannon, 1948). For example, if we want to know the size of a lake, we can measure its area by means of an electronic map. The smaller the scale of measurement, the more accurate the result of measurement. If the scale becomes smaller and smaller, and the measurement results converges rapidly to a certain value, then we can say that the lake area has a characteristic scale. The characteristic scale can be represented by the radius length of the lake's equivalent circle (Taylor, 1983).

In urban geography, urban land use area can be used to reflect the extent of space filling. However, if we try to measure the total area of land use of a city, the process will become very complicated. First, it is hard to identify a clear urban boundary line. Second, the patches of urban land use on remoting sense images are too random, fragmented, and irregular to be exactly caught. Although the scale of measurement becomes smaller and smaller, the measurement results will never converge. Finally, the measurement process reaches the limit allowed by image resolution and is cut off. So, in order to characterize the land-use filling degree of urban space, we must find a new approach.

According to fractal theory, urban area can be replaced by a scaling exponent, that is, fractal dimension. By the double logarithmic linear relation between scales (say, linear sizes of boxes) and corresponding measures (say, numbers of nonempty boxes), we can estimate a slope on a log-log plot (Chen *et al*, 2017; Feng and Chen, 2010). The value of the slope indicates the fractal dimension, and the parameter can reflect the space filling extent of urban land use.

It can be seen that there is a symmetry and duality relation between Euclidean geometry and fractal geometry. It is just Euclidean geometric measurement that leads to fractal dimensions. For a Euclidean geometry body, the dimension is known: 0 dimension for points, 1 dimension for lines, 2 dimensions for surfaces, and 3 dimensions for bodies. But generally speaking, without measurement, we cannot know the length, or area, or the volume. In contrast, for a fractal object, the length, or area, or volume, is known in principle: if the topological dimension is 1, the length is infinity; if the topological dimension is 0, the length, or area, or volume, is 0. In theory, the Lebesgue measure of a real fractal object is zero. This suggests that a fractal defined in a 2-dimensional embedding space have no area, and a fractal defined in a 3-dimensional embedding space has no volume. However, without measurement and calculation, we cannot know its fractal dimension value. In the process of fractal measurement, common scales are replaced by scaling, and conventional measures such as length, area, and volume are replaced by fractal dimension. Where urban form is concerned, both area and fractal dimension can reflect space filling extent. If the area measurement of a city fails, we will use fractal dimension to replace the area to characterize it space filling.

## 2.2 Uniqueness and diversity of fractal dimension

A geometric phenomenon has only one dimension value, which is of Euclidean dimension or fractal dimension. In other words, for a given aspect of a geometric object, the dimension value is uniquely determined. For example, for a circle or a square, it has two aspects: area and perimeter. The dimension for area is 2, and the dimension for boundary line is 1. Where a fractal is concerned, thing seems to be more complicated, but can be made clear. For a regular fractal defined in 1-dimension embedding space and its topological dimension is 0 (e.g., Cantor set), or for a fractal line defined in 2-dimension embedding space but its topological dimension equals 1 (e.g., Koch curve, Peano curve), it has only one aspect and the fractal dimension is uniquely determined in theory. For a fractal object defined in 2-dimension embedding space and topological dimension is 0 (e.g., Box growing fractal, Sierpinski gasket), it has two aspects corresponding to two different fractal dimensions. As a special example, Vicsek fractal (box growing fractal) has fractal area (aspect 1) and fractal boundary (aspect 2), both the two fractal dimension values are $D=\ln(5)/\ln(3)=1.465$ (Vicsek, 1989); Sierpinski gasket also has aspects, the fractal dimension of fractal area is $D=\ln(3)/\ln(2)=1.585$ (Mandelbrot, 1982), while the fractal dimension of fractal boundary is $D=\ln(5)/\ln(2)=2.3219$ (Chen, 2010). Please note that the similarity dimension and radial dimension can exceed the dimension value of its embedding space. The Koch lake is a special case, which also has two aspects: area and perimeter. The dimension for area is 2, and the dimension for boundary curve is 1.2619 (Mandelbrot, 1982). The perimeter of the Koch lake is a fractal line (topological dimension is 1), but the area within the perimeter is a Euclidean plane (topological dimension is 2).

Fractal studies on cities are neither pure mathematical processes nor absolute true portrayal of cities. Just like any other scientific research, urban studies based on fractal geometry are involved with three worlds, that is, *real world*, *mathematical world*, and *computational world* (Casti, 1996). The mathematical world (objective world) is always linked to the real world (objective world) by the computational world (subjective world) (Figure 1). Random fractals in the *real world* are more complex than the regular fractals in the *mathematical world*. For a fractal city defined in the 2-dimension space, if we examine the land-use pattern, it will be involved with two aspects, *urban area* and *urban boundary*. The fractal dimension of urban form can be estimated by box-counting method (Benguigui *et al*, 2000; Chen and Wang, 2013; Feng and Chen, 2010; Jiang and Liu, 2012; Shen, 2002; Sun and Southworth, 2013), sandbox method (Ariza-Villaverde *et al*, 2013), area-radius scaling (cluster growing) method (Batty and Longley, 1994; Frankhauser, 1998; Jiang and Zhou, 2006; Sambrook and Voss, 2001; White and Engelen, 1993), density-radius scaling method (Batty, 1991), length-area scaling (Rodin and Rodina, 2000), correlation and dilation methods (De Keersmaecker *et al*, 2013), wave-spectrum scaling method (Chen, 2013), and so on; the fractal dimension of boundary line can be directly estimated by walking-divider method (Longley and Batty, 1989a), or indirectly estimated by perimeter-area scaling method (Batty and Longley, 1988; Benguigui *et al*, 2006; Longley and Batty, 1989b), and so on. Unfortunately, in empirical studies on city fractals, the process is very complicated. Differing from the field investigation in the *real world* and logic reasoning in the *mathematical world*, the fractal dimension estimation is conducted in the *computational world*. However, the subjectivity of fractal dimension measurement and calculation is not the fatal problem, the key lies in the fractal properties of cities, and this will be discussed in next section.

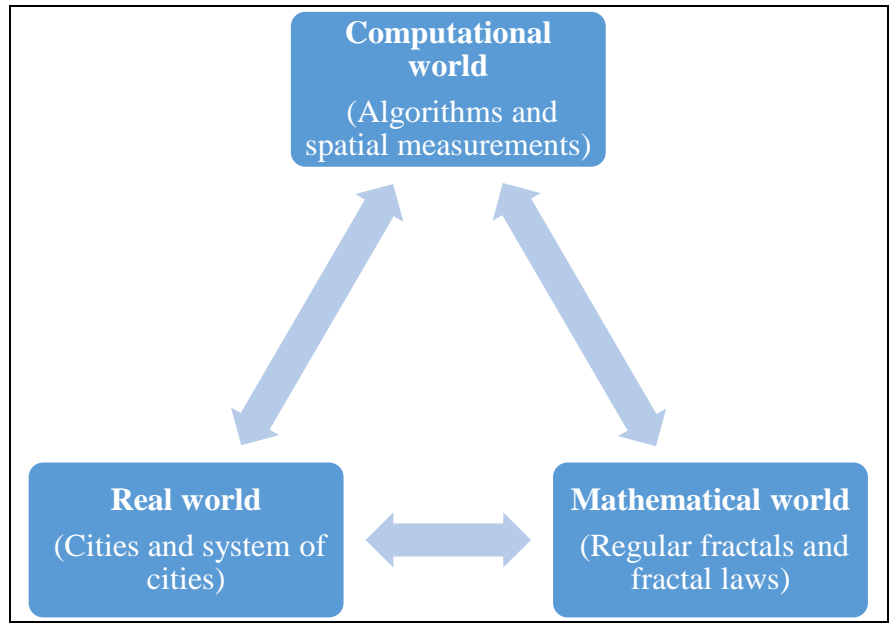

**Figure 1** Fractal studies on cities defined in three worlds: real world, mathematical world, and computational world

**Note**: The diagram is constructed by referring to the work of Casti (1996). The relationships between the three worlds and fractal concepts are illustrated in this schema.

## 2.3 Varied fractal dimension calculation methods

In order to calculate fractal dimension, fractal scientists propose a number of methods. The most of these methods are generic in different fields, and they were summarized and sorted by fractal experts years ago (Takayasu, 1990). In urban studies, fractal dimension estimation methods were once researched, developed, and sorted by urban theoreticians such as Michael Batty, Frankhauser, and Paul Longley (Batty and Longley, 1994; Frankhauser, 1998). Generally speaking, different methods have different uses. But sometimes, we can use different methods to estimate the same fractal dimension. For example, four types of methods can be used to estimate the fractal dimension geographical boundaries (Batty and Longley, 1994; Longley and Batty, 1989b). Now, various fractal dimension methods have been introduced into fractal city studies.

**First, different methods can be employed to compute the fractal dimensions for different aspects of a city or a system of cities.** Based on time series of urban evolution, the fractal

dimensions of dynamic processes can be estimated; based on spatial datasets, the fractal dimensions of spatial structure of cities can be estimated; based on sectional data of cities, the fractal dimensions of rank-size distributions and hierarchies can be estimated (Table 1). The datasets of time series, space series, and hierarchy series can be used to make different types of spatial analyses and correlation analyses for urban studies.

Table 1 Methods of urban fractal dimension estimation based on time series, spatial structure and hierarchical structure

| Object | Method | Fractal dimension |
| --- | --- | --- |
| **Time series (process)** | Power spectrum | Self-affine and self-similar dimension |
| | Reconstructing phase space | Correlation dimension |
| | Elasticity relation | Similarity dimension |
| | …… | …… |
| **Spatial structure, texture, and distribution (pattern)** | Box counting method | Self-similar dimension |
| | Sandbox | Self-similar dimension |
| | Radius scaling (cluster growing) | Self-similar dimension |
| | Wave spectrum | Self-affine and self-similar dimension |
| | Walking-divider method | Self-similar dimension |
| | Perimeter-area scaling | Self-similar dimension |
| | …… | …… |
| **Hierarchical structure (cascade), Scale-free network** | Size distribution | Self-similar dimension |
| | Hierarchical scaling | Self-similar dimension |
| | Allometric scaling | Self-similar dimension |
| | Renormalization | Self-similar dimension |
| | …… | …… |

**Second, different methods can be utilized to calculate different types of urban fractal dimension.** In the most majority of cases, we compute the self-similar fractal dimensions of urban form or urban systems. But sometimes, we are concerned about the self-affine fractal dimension of urban growth (Chen and Lin, 2009). Self-similar fractal dimension and self-affine fractal dimension represent two different but related fractal parameters (Table 2). Self-similar growth is isotropic growth, while self-affine growth indicates anisotropic growth (Vicsek, 1989). Generally speaking, the density distribution of urban transport network takes on self-similar growth with isotropy, while urban land-use expansion takes on self-affine growth with anisotropy.

**Table 2 Fractal dimension estimation methods for self-similar patterns and self-affine processes of cities**

| Fractality | Aspect | Method |
|---|---|---|
| **Self-similarity** | Area/Point (spatial structure) | Box counting method |
| | | Prism counting method |
| | | Area-radius scaling (cluster growing) |
| | | Sandbox method |
| | | Wave spectrum analysis |
| | | …… |
| | Boundary/Line (spatial texture) | Walking-divider method |
| | | Perimeter-area scaling |
| | | …… |
| | Network | Renormalization |
| | | …… |
| **Self-affinity** | Area/Line | Fractional Brownian Motion (FBM) |
| | | Wave spectrum |
| | | …… |

**Third, a variety of methods can be used to compute the same fractal dimension of cities.** The solution to a problem is not the only one. For each aspect of city fractal, more than one approaches can be used to estimate a fractal dimension (Table 3). For example, we can use walking-divider method, perimeter-area scaling method, and box-counting method to estimate the fractal dimension of urban boundary (Longley and Batty, 1989a; Batty and Longley, 1994; Longley and Batty, 1989b; Wang *et al*, 2005). In theory, the three methods are equivalent to each other. An aspect of a fractal object has only one fractal dimension. However, in practice, the calculation results from different methods are not always consistent with each other. This gives rise to a number of problems about fractal dimension calculation of cities.

**Table 3 Direct and indirect fractal dimension estimation methods for cities**

| Property | Method (type) | Method (subtype) |
|---|---|---|
| **Direct** | Box counting | Common box, prism box, sandbox |
| | Radius scaling (cluster growing) | Area-radius scaling, number-radius scaling, density-radius scaling, radius of gyration |
| | Walking-divider | Various step length processes |
| | …… | …… |
| **Indirect** | Spectral analysis | Wave spectrum, power spectrum |

|  | Geometric measure relation | Allometric scaling, perimeter-area scaling, length-area scaling, elasticity relation |
|---|---|---|
|  | Fractional Brownian Motion | (mainly for self-affine process) |
|  | …… | …… |

## 3. Solutions to fractal dimension estimation problems

### 3.1 A dilemma of fractal dimension estimation

To make or use a mathematical model, we must find an effective algorithm and approach to determine its parameter values. The algorithms include the ordinary least squares (OLS), maximum likelihood estimation (MLE), and major axis method (MAM). A number of measurement approaches, as displayed above, are proposed in literature to estimate fractal dimension values (Tables 1, 2 and 3). Generally speaking, different methods are applied to different directions (different aspects or properties). For example, walking divider method can be used to estimate the fractal dimension of urban boundary dimension rather than urban area, power spectrum is used to research the urban evolution based on time series rather than urban form based on spatial data, fractional Brownian motion (FBM) is used to estimate self-affine record dimension rather than self-similar trail dimension, sandbox method, clustering growing and wave-spectrum are used to calculate the radial dimension for characterizing urban growth, box-counting method is used to compute fractal dimensions for describing spatial structure and texture of urban morphology, and so on. Sometimes, several different methods can be applied to the same aspect of cities. For example, box-counting method, area-radius scaling method, sandbox method, and wave spectrum analysis based on density-radius scaling can be employed to estimate the fractal dimension of urbanized area. In theory, as indicated above, a fractal aspect have only one fractal dimension value. But unfortunately, in empirical studies, different methods often result in different fractal dimension estimation values, and in many cases, the numerical differences are statistically significant and cannot be ignored in a spatial analysis. Even for a given method, a fractal dimension value often depends on the size and central location of the study area defined by a researcher. This is involved with the uncertainty of fractal dimension calculation, which puzzles many fractal scientists.

A simple prototype is helpful for understanding complex phenomena in scientific research. In order to study the atomic structure, physicists first explored the structure of the simplest atom, hydrogen atom; In order to study the structure of viruses, biologists first concentrated on exploring the structure of simple virus, bacteriophages. Simple prototypes often form the beginning of theoretical analysis. To reveal the root of the problem of uncertainty in fractal dimension calculation, we can examine two regular fractals, including monofractal and multifractal patterns. All these regular fractals reflect prefractal structure because we can never look the real fractal patterns. The real fractals in geometry are just like the high-dimensional spaces in linear algebra, which can be imagined but cannot be observed. All of the fractal images we encounter in books or articles represent prefractals rather than real fractals (Addison, 1997). The difference between real fractals and prefractals is as below: A real fractal has infinite levels, but a prefractal is a limited hierarchy; therefore, the Lebesgue measure of a real fractal equals 0, but the Lebesgue measure of a prefractal is not equal to 0. For a given aspect (say, area, or boundary) of a regular monofractal object, we can applied different methods to its prefractal structure to determine its fractal dimension. Different methods lead to the same result, which represents the real fractal dimension value. However, for a multifractal object, the real fractal dimension cannot be computed by applying some method to its prefractal pattern. We can only obtain comparable parameters rather than real fractal dimension for multifractal systems.

By analyzing the regular fractal objects, we can gain new insight into fractal structure and fractal dimension measurement. First of all, let's see a simple regular growing fractal, which is employed to model urban growth in literature (Batty and Longley, 1994; Chen, 2012; Frankhauser, 1998; Longley *et al*, 1991; White and Engelen, 1993). This fractal was proposed by Jullien and Botet (1987) and became well known due to the work of Vicsek (1989), and it is also termed Vicsek's figure or box fractal (Figure 2). Three approaches can be applied to its prefractal pattern, including box-counting method, sandbox method, and cluster growing scaling method. The third approached can be divided into two equivalent methods: area (number)-radius scaling and density-radius scaling. According to its regular composition, we can obtain the datasets comprising the first 10 steps (Table 4). Based on box-counting method, sandbox method, and area-radius scaling method, the scaling exponent is just its fractal dimension, and the value is $D=\ln(5)/\ln(3)=1.465$. Based on density-radius scaling method, the scaling exponent is $a=2-D=0.535$, and thus the fractal dimension is $D=2-0.535=1.465$. This value is exactly the real fractal dimension of this fractal object.

**Table 4** Box-counting method, sandbox method, and cluster radius scaling methods for fractal dimension of a regular monofractal growing fractal

| Level | Box-counting method | | Sandbox method | | Cluster growing and radius scaling | | |
|---|---|---|---|---|---|---|---|
| $m$ | Box side length $r$ | Box number $N(r)$ | Sandbox side length $L$ | Box number $N(L)$ | Radius $R$ | Fractal unit number $N(R)$ | Density $\rho(R)$ |
| **0** | 1.0000 | 1 | 1 | 1 | 0.7071 | 1 | 1.0000 |
| **1** | 0.3333 | 5 | 3 | 5 | 2.1213 | 5 | 0.5556 |
| **2** | 0.1111 | 25 | 9 | 25 | 6.3640 | 25 | 0.3086 |
| **3** | 0.0370 | 125 | 27 | 125 | 19.0919 | 125 | 0.1715 |
| **4** | 0.0123 | 625 | 81 | 625 | 57.2756 | 625 | 0.0953 |
| **5** | 0.0041 | 3125 | 243 | 3125 | 171.8269 | 3125 | 0.0529 |
| **6** | 0.0014 | 15625 | 729 | 15625 | 515.4808 | 15625 | 0.0294 |
| **7** | 0.0005 | 78125 | 2187 | 78125 | 1546.4425 | 78125 | 0.0163 |
| **8** | 0.0002 | 390625 | 6561 | 390625 | 4639.3276 | 390625 | 0.0091 |
| **9** | 0.0001 | 1953125 | 19683 | 1953125 | 13917.9828 | 1953125 | 0.0050 |
| **…** | … | … | … | … | … | … | … |

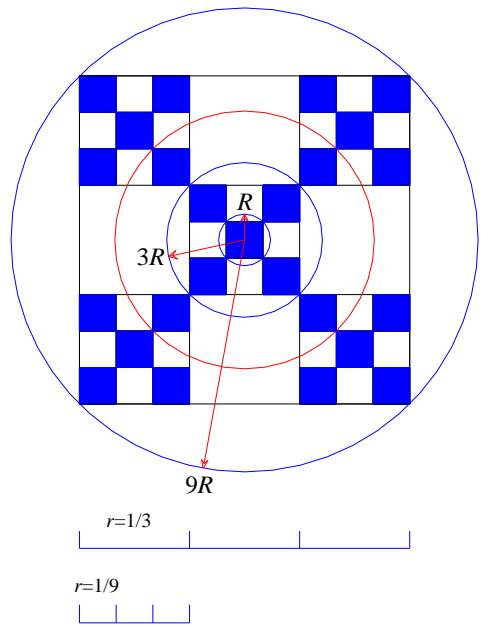

**Figure 2** Three approaches to estimating fractal dimension of a regular fractal (the first 3 steps)

**Note**: The schematic diagram of measurement method is drawn by referring to the work of Batty and Longley (1994). Sandbox method, radius-number scaling, and box-counting method can be employed to calculate the fractal dimension of this growing fractal.

Further, let's examine a regular growing multifractal object, which reflect the pattern of spatial heterogeneity. This fractal is presented by Vicsek (1989). The first three steps represent a prefractal process (Figure 3). Box-counting method can be used to calculate its global dimension. Step 1: fractal dimension $D=0$ (for a point, the fractal dimension can be obtained by L'Hospital's rule). Step 2: box dimension $D=- \ln(17)/\ln(1/5)=1.7604$. Step 3: box dimension $D= -\ln(289)/\ln(1/25)=1.7604$. If we apply sandbox method to the figure in the third step, the fractal dimension is also $D=1.7604$. However, two problems can be found by careful investigation. **First, different fractal units bear different fractal dimension values.** One of basic properties of fractals, including monofractals and multifractals, is entropy conservation: different fractal units at a given level has the same Shannon entropy value. However, the fractal dimension does not comply with a conservation law. In fact, for a multifractal system, different parts have different local fractal dimensions. For example, for the second level of the third step, the five parts have two fractal dimension values. The central part, box dimension is $D=\ln(1/17)/\ln(1/5)=1.7604$; the other four parts, box dimension is $D=\ln(4/17)/\ln(2/5)=1.5791$. **Second, the parameter value estimated by the box-counting method and the sandbox method is not equal to its real dimension value.** In theory, the calculated values represent the capacity dimension of this multifractals, i.e., $D_0=1.7604$. The regular multifractal structure can be modeled by a transcendental equation based on probabilities and the corresponding scales. Where the third step is concerned, the multifractal transcendental equation can be constructed as follows

$$\sum_{i=1}^{5} P_i^q r_i^{(1-q)D_q} = (\frac{1}{17})^q (\frac{1}{5})^{(1-q)D_q} + 4(\frac{4}{17})^q (\frac{2}{5})^{(1-q)D_q} = 1. \tag{1}$$

Using Matlab to find its numerical solutions, we can obtain its multifractal parameter values (Table 5). The results show that the real capacity dimension is about $D=1.5995 < D= 1.7604$. The capacity dimension based on box-counting method and sandbox method is in fact the maximum dimension, that is $D_{-\infty}=1.7604$. The capacity ($D_0$) is the maximum value of the local dimension, while the maximum dimension ($D_{-\infty}$) is the upper range value of global dimension.

Table 5 Four sets of fractal parameters of a regular growing multifractals (typical values)

| Moment order $q$ | Global parameters | | Local parameters | |
|---|---|---|---|---|
| | Global dimension $D_q$ | Mass exponent $\tau_q$ | Singularity exponent $\alpha(q)$ | Local fractal dimension $f(\alpha(q))$ |
| **-100** | 1.7429 | -176.0374 | 1.7604 | 0.0000 |
| **-10** | 1.6404 | -18.0440 | 1.6933 | 1.1107 |
| **-2** | 1.6054 | -4.8161 | 1.6153 | 1.5855 |
| **-1** | 1.6022 | -3.2044 | 1.6081 | 1.5963 |
| **0** | 1.5995 | -1.5995 | 1.6020 | 1.5995 |
| **1** | 1.5970 | 0.0000 | 1.5970 | 1.5970 |
| **2** | 1.5949 | 1.5949 | 1.5930 | 1.5910 |
| **10** | 1.5859 | 14.2730 | 1.5806 | 1.5330 |
| **100** | 1.5798 | 156.3975 | 1.5791 | 1.5129 |

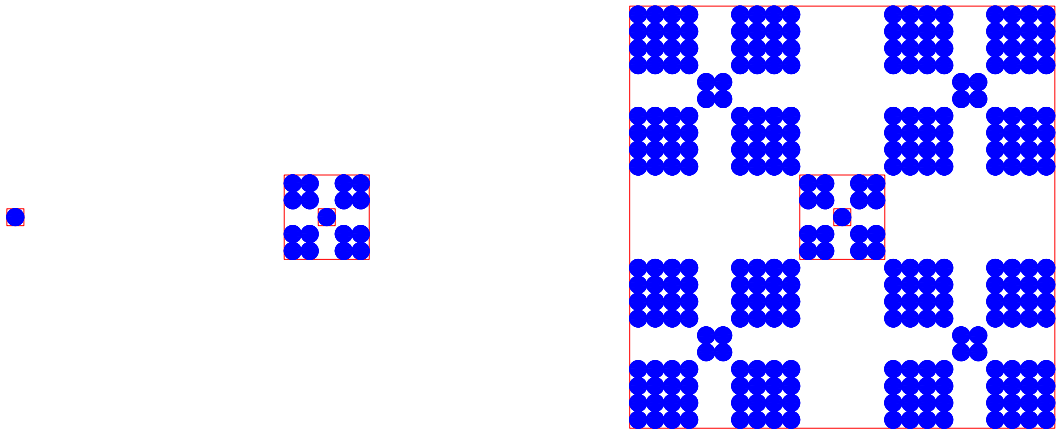

**Figure 3** The prefractal structure of a regular growing multifractals (the first three steps)

**Note**: The fractal pattern is adapted by referring to the work of Vicsek (1989). This fractal can be used to model multifractal growth of cities (Chen and Wang, 2013; Wang and Li, 1995).

Now, a basic judgment can be reached as follows. For a regular monofractal (Figure 2), the real fractal dimension value can be calculated by the prefractal structure. However, for a regular multifractal (Figure 3), the real fractal dimension values cannot be obtained by applying some method such as box-counting method to its prefractal structure. For a random multifractal system, we cannot construct its multi-scaling transcendental equation such as equation (1). As a result, we will never know the real fractal parameter values. We can only estimate a set of comparable parameter values to replace the real values. In the real world, fractal cities have two properties. First, they are random multifractals rather the regular monofractals or regular multifractals; second, they only develop prefractal structure rather real fractal structure. What is more, the prefractal structure of multifractals are always mixed up with self-affine processes, fat fractal components, or even fractal complements. In this case, the processes of measurements and analyses become very complicated relative to the regular fractals.

## 3.2 The reasons for the divergence of calculation results

In urban studies, fractal dimension value is always influenced by selection of method and definition of study area. Fractal dimension estimation depends on method, and this is indeed a problem. However, study area (size, location) influence fractal dimension values, but this is not a problem. The concrete reasons are as follows (Table 6). **(1) Prefractal is the main reason for the influence of method on the fractal dimension values.** For a real fractal, its scaling range is infinite; for a prefractal, its scaling range is limited to certain scales. The precondition of accurate calculation of fractal dimension for a random fractal is that the scale of measurement is close to infinitesimal in theory. For a given aspect of a given random fractal, if measurement scale become smaller and smaller, different methods will lead to the similar fractal dimension values. However, for a pre-fractal, the linear size of measurement scale is limited to its lower bound of scaling range and cannot approach infinitely small scales. **(2) Multifractal structure is the main reason for the influence of the size and central location of the study area on the fractal dimension values.** The fractals in the real world are all random multifractal rather than monofractals. A monofractal object has only one scaling process, while multifractals have more than one scaling process. For a monofractal object, capacity dimension is equal to information dimension and correlation dimension, and global dimension is equal to local dimension. However, for a random multifractal, capacity dimension is greater than information dimension, and information dimension is greater than correlation dimension, and so on. Different parts of a random multifractal object have different local fractal dimension, and different size of study area yield different fractal parameters. Thus, if we define different study area for a multifractal city, the results of fractal dimension estimation will be different. **(3) Self-affine fractal process is a reason associated with the influence of both method and study area on fractal dimension estimation.** Self-similar growth indicates isotropy, and measurement direction does not influence fractal dimension estimation, while self-affine growth implies anisotropy, and different measurement directions lead to different fractal dimension. Especially, self-affine growth causes area-radius scaling to break, and form what is called bi-fractal pattern in a log-log plot. The essence of bi-fractals rests with self-affine development and growth of fractal systems, which can be illustrated by testing the regular self-affine fractals.

**Table 6 Three significant properties of city fractals: prefractal structure, multifractal form, and self-affine growth**

| City fractal | Theoretical problem | Practical problem |
|---|---|---|
| **Random prefractal** | The range of measurement is limited. The topological dimension is easily misunderstood, and this leads to misunderstanding on scaling range. | Image resolution influence the identification of patterns, which in turn influence fractal dimension estimation. |
| **Random multifractal** | Different moment order $q$ lead to different global fractal dimensions, different parts have different local fractal dimensions. | The scope of study area and the angle of view influence the multifractal parameter spectrums. |
| **Random self-affine fractal** | Anisotropic growth lead to different fractal dimension values in different directions. | It is hard to estimate fractal dimension using radius scaling method. |

## 3.3 Solutions to problems

The problems on the uncertainty of fractal dimension estimation cannot be solved once and for all. Different problems should be treated differently, and different types of problems need different types of solutions (Table 7). The main ideas to solve the problem are two: one is to find a proper method for a special studies on fractal cities, and the other is to replace the real fractal parameters with the comparable fractal parameters. Based on the above explanation, the possible solutions to the problems of fractal dimension estimation of cities can be presented as follows.

**First, the solutions to the problem of method dependence of fractal dimension estimation**. The results of model parameter estimation depend on methods, this phenomenon is not only in the field of fractal research. It is hard to find good solutions to this kind of problem. In fact, for a random system or based on random variables, it is unlikely to find the true parameter values for its mathematical models. Scientists then look for comparable parameter values instead of real parameter values (*Faute de mieux*). In urban studies on fractals, two approaches are employed to deal with this kind of problems. One approach is to replace the real fractal values by comparable fractal dimension values which are based on the same criterion for different times, spaces, levels, and scales; the other approach is find the most suitable method for specific research objectives. As indicated above, different methods have different merits and can be applied to different aspects and directions of urban studies (Tables 1, 2, and 3). If we measure the fractal dimension of geographical fractal lines such as urban boundaries, the advisable methods are walking divider method and area-perimeter scaling method (Batty and Longley, 1994; Longley and Batty, 1989); if we research complex patterns and spatial structure, we can utilized box-counting method (Benguigui *et al*, 2000; Chen and Wang, 2013; Feng and Chen, 2010; Shen, 2002); if we explore the dynamic process of isotropic urban growth, we should adopt sandbox method, area-radius scaling, density-radius scaling (Ariza-Villaverde *et al*, 2013; Batty and Longley, 1994); if we examine anisotropic urban growth, we should adopt wave-spectral analysis based on density-radius scaling (Chen, 2010; Chen, 2013). Generally speaking, in order to estimate the fractal dimension of an urban boundary, we often regard urban area as Euclidean surface with an integral dimension *d*=2 (Batty and Longley, 1994). Thus we can make use of area-perimeter scaling method (Feder, 1988). However, if we try to compare the fractal dimension value of urban form with that of urban boundary for the same city, we should adopt box-counting method, which can give comparable fractal dimension for both boundary line and urban pattern within this boundary.

**Second, the solutions to the problems of study are scope dependence of fractal dimension calculation**. A random prefractal object has a limited scaling range, in which fractal property appears. As shown above, for a given aspect (area or boundary) of a regular momofractal, its fractal dimension value is unique, and the real fractal dimension can be calculated through its prefractal structure. However, for a regular multifractal object, different parts have different local fractal dimension values, and the real fractal dimension values cannot be computed by its prefractal structure. No regular monofractal can be found in the real world. A real city is a random multifractal system with prefractal structure. It is impossible to identify the boundaries of different fractal units. Different sizes of study area processes different global fractal dimensions, and different parts have different local fractal dimensions. Consequently, based on different scope (size and central location) of a study area, different fractal parameter values will be worked out. We never know the sets of real fractal dimension values. In this case, we can use relative comparable parameter values instead of absolute real parameter values. In particular, we can employ multifractal dimension spectrums to make spatial analyses for cities and systems of cities. To obtain comparable fractal dimension, we must define comparable study area. Where urban form and box-counting method are concerned, the procedure is as below: (1) using a proper method and the concept from characteristic scales to define objective urban boundaries; (2) defining a measure area based on certain direction for the urban envelope; (3) use the measure area as the maximum box for box counting. In the specific research, the methods should be adjusted according to specific problems and research objectives.

Table 7 The possible directions of solving problems in fractal dimension estimation

| Factor | | Reason | Mechanism | Influence | Solution |
| --- | --- | --- | --- | --- | --- |
| **Method** | | Pre-fractal | Scaling range | Analytical conclusions | Select the most suitable method |
| **Study area** | Size of study area | Pre-multifractals | Multi-scaling pattern and range | Analytical objects | Define a comparable scope |
| | Place of study area | Pre-multifractals | Multi-scaling process and range | Analytical objects | Define a comparable location |

## 4. Questions and discussion

In scientific research, if we cannot obtain absolute measurement based on certain values, we should try to find the relative measurement based on comparable values. If we only focus on the fractal studies on cities, the uncertainty of fractal dimension estimation is a problem; however, if we look at the entire system of scientific methodology, then this kind of uncertainty is not a problem. In fact, the uncertainty of model parameter estimation is a common phenomenon in scientific research. Mathematical models and quantitative analyses can be divided into two types: one is based on characteristic scales, and the other is based on scaling. The traditional mathematical tools are mainly based on characteristic scales, while fractal studies are mainly based on scaling. In conventional mathematical modeling processes, the parameter estimation relies heavily on computational methods. It is impossible to evaluate the real parameters for the great majority of mathematical models by empirical analysis. A number of examples are listed as follows (Table 8). (1) For the simplest linear regression model, a number of algorithms such as least squares method, maximum likelihood method, major axis method, and reduced major axis method can be employed to estimate the regression coefficients, and different methods lead to different results. Moreover, sample size and variable dimension also influence the constant and regression coefficients. (2) For factor analysis, the calculation results depend on the methods of factor extraction and factor rotation, and there are various methods for factor extraction (e.g., orthogonal transformation, maximum likelihood) and rotation (e.g., Quartimax, Varimax). What is more, the starting point of factor analysis can be correlation coefficient matrix or covariance matrix, and different starting points lead to different numerical results. (3) For hierarchical cluster, the final output depends on the methods of cluster and measure, and there are various method for cluster (e.g., between-groups linkage, within-groups linkage) and measure (e.g., Euclidean distance, Pearson correlation). Different measure methods lead to different proximity matrixes, and different cluster methods based on different proximity matrixes lead to different final results. Moreover, the value transform methods (e.g., standardization, normalization, etc.) influence cluster analysis. (4) For auto-regression analysis based on time series, there are various methods for parameter estimation; for spatial autocorrelation analysis, different impedance functions lead to different contiguity matrixes, which in turn lead to different Moran's $I$, Geary's $C$, Getis's $G$, and so on.

**Table 8 Diversity of methods for estimating model parameters or finding solutions to problems**

| Type | Model | Methodology | |
|---|---|---|---|
| | | Category | Approach |
| **Characteristic Scale** | Regression analysis | Algorithm | Least squares, Maximum likelihood, Major axis, Reduced major axis, … |
| | Factor | Extraction | Principal components, Unweighted least squares, Generalized least squares, Maximum likelihood, Principal axis factoring, Alpha factoring, Image factoring, … |
| | | Rotation | None, Quartimax, Varimax, Equamax, Promax, Direct oblimin, … |
| | | Analytical base | Correlation matrix, covariance matrix |
| | Hierarchical cluster | Cluster | Between-groups linkage, Within-groups linkage, Nearest neighbor, Furthest neighbor, Centroid clustering, Median clustering, Ward's method, … |
| | | Measure | Euclidean distance, squared Euclidean distance, cosine, Pearson correlation, Chebychev distance, Block distance, Mahalanobis distance, Minkowski distance, varied customized distance, |
| | | Value transform | None, standardization (Z scores), range standardization (range -1 to 1), range normalization (range 0 to 1), maximum |

|  |  |  | magnitude of 1, mean of 1, standard deviation of 1,… |
|  | Auto-regression | Algorithm | Exact maximum-likelihood, Cochrane-Orcutt, Prais-Winsten, Least squares, … |
|  | Spatial autocorrelation | Measurement | Moran's *I*, Geary's *C*, Getis' *G*, Ripley's *K*, … |
|  |  | Calculation | Conventional formula, Three-step calculation, Matrix scaling, Standard deviation, Least square, … |
|  |  | Contiguity matrix | Power function, exponential function, step function, … |
| **Scaling** | Fractals | Algorithm | Least squares, Maximum likelihood, Major axis, Reduced major axis, … |
|  |  | Measurement | Box-counting, sandbox, radius scaling, radius of gyration, walking divider, geometric measure relation, spectral analysis, distribution function, … |

The uncertainty of fractal dimension calculation is associated with the uncertainty of scaling exponent estimation. Scaling is one of basic properties of fractals. If and only if a scaling phenomenon satisfies three conditions, it can be regarded as a fractal set. The conditions include scaling law (scale invariance), fractal dimension (Hausdorff dimension is greater than its topological dimension), and entropy conservation (the Shannon entropy of each fractal units is a constant) (Chen, 2016). However, many natural and social complex systems follows scaling law, but have no fractal dimension and do not meet the entropy conservation condition. These types of complex systems cannot be effectively modeled by traditional mathematical methods. In this case, we can use scaling exponents to characterize the complex systems. In recent years, scaling has become a hot topic in urban studies, and a number of interesting research results emerged (Bettencourt, 2013; Bettencourt *et al*, 2007; Bettencourt *et al*, 2010; Lobo *et al*, 2013; Jiang and Jia, 2011; Jiang and Liu, 2012; Pumain *et al*, 2006). Among various urban scaling, the most frequently appearance is the allometric scaling. However, in empirical studies, it is difficult to obtain stable scaling exponent values. Algorithms, study area, datasets, scaling ranges, and so on, influence the results of scaling exponent estimation (Chen, 2017). A recent discovery is that the scaling exponent values of the allometric relation between patents and city sizes depend on the population size cut-offs (Arcaute *et al*, 2015); Another meaningful discovery is that the scaling exponent values of the allometric relation between urban CO$_2$ emissions and city population sizes depends on the definition of urban area (Louf and Barthelemy, 2014a; Louf and Barthelemy, 2014b). A scaling exponent is often directly or indirectly related to fractal dimension. The allometric scaling exponent is actually the ratio of one fractal dimension to another fractal dimension (Chen, 2017). In this sense, the uncertainty of fractal dimension computation account for the uncertainty of scaling exponent estimation of cities.

As indicated above, fractal dimension calculation is implemented in the computational world rather than in the mathematical world. Cities appearing in the real world are objective, but it is hard to reveal the deep structure and the complicated relationships between causes and effects hidden behind urban world. Regular fractals, fractal laws, and strict logic reasoning defined in the mathematical world are also objective, but the graceful mathematical processes are not consistent with the real systems. We can use Koch snowflake to model central place system of human settlements, but the real central place networks differ from the ideal Koch snowflake pattern. We can employ the diffusion-limited aggregation models to simulate urban growth and form, but real urban evolution differs from DLA process. So, Albert Einstein once said, "I don't believe in mathematics." He observed, "As far as the laws of mathematics refer to reality, they are not certain, and as far as they are certain, they do not refer to reality." In fact, as a pure theoretical physicist, Einstein ignored an important linkage, which represents a logic bridge between mathematical world (e.g., theoretical models and laws) and real world (e.g., urban growth and form). The bridge coming between reality and mathematics is what is called computational world, which is a subjective world to some extent (Figure 1). Spatial measurements, data processing, algorithms, and so on, are all defined in the computational world. The ways of measurements and data extraction as well as choices of algorithms and models varies from person to person. Therefore, for the urban form of a same city, the fractal dimension estimation results may be different from one another significantly. The more experienced a scholar is, the better the process of fractal dimension calculation is handled. However, no matter how hard we try, we can never obtain the real values or absolutely exact values for the fractal parameters of a city. As indicated above, the best results that we can gain in a study are a set of comparable fractal dimension values for different times, places, levels, or scales.

The three worlds are related to three types of studies about fractal cities. According to the theory of systems analysis, academic research falls into three categories: *behavioral research*, *normative research*, and *values research* (Krone, 1980). Behavioral research on fractal cities are a type of positive studies, which correspond to the real world; normative research on fractal cities are pure theoretical studies, which correspond to the mathematical world. On the one hand, fractal geometry is a powerful tool for exploring nonlinear processes, irregular patterns, and scale-free distributions (spatial distributions and probability distributions), and can be used to bright to light the evolution process and spatial pattern of cities. On the other, fractals represent the optimum structure in nature and society. A fractal object can occupy its space in the most efficient way. In this sense, fractal geometry can be devoted to finding the optimized structure of urban systems or construct ideal models for urban spatial analyses. The two types of research can be linked by the values research. For fractal cities, the values research is to develop a set of evaluation indexes, by which we can judge the pros and cons of the development of a city in the past and at present (for behavioral research) and its direction of optimal design in the future (for normative research).

The emergence of fractal geometry represents a discovery of new form of symmetry, i.e. scaling symmetry. The basic property of a fractal is its invariance under contraction or dilation (1989). Because of the scaling symmetry, it is impossible to find certain length, area, volume, and number for a scale-free system. In this case, we can use a scaling exponent to replace the common measures. One of the basic scaling exponents is fractal dimension. As a matter of fact, there must be some symmetry when there is immeasurable quantity (Lee, 1988). As indicated above, fractal concept came from the immeasurable length of the cost of Britain (Mandelbrot, 1967). Today, fractal dimension seems to be another immeasurable quantity. Although we found various factors that affect fractal dimension measurements, there is no exclusion of the possibility that a kind of super symmetry is hidden behind the scaling symmetry (Chen, 2008). What is more, spatial autocorrelation of urban patterns influence of spatial measurement results. Reliable measurements depend on no spatial autocorrelation. These problems remains to be explored in future studies. All in all, we cannot give up eating for fear of choking, and cannot give up fractal geometry in urban studies because of the uncertainty of fractal dimension estimation. The application value of a measure or parameter value lies in comparability rather than reality or accuracy. It's like a small scale map of a country or the world. A map is a typical model, and the mapping is a typical process of model building (Holland, 1998). When we map the geographical things on the three-dimensional spherical surface to the two-dimensional plane, in any case, we will encounter the projection deformation, which results in the distortion of the spatial pattern on the map. However, the maps are very useful in everyday life and geographical research.

# 5. Conclusions

There are various approaches to fractal dimension estimation, and the great majority of them can be adopted to research fractal cities. Generally speaking, different methods are suitable for different directions of urban studies. Sometimes, several different methods can be applied to the same aspect of fractal dimension estimation, but the results are different from each other significantly. What is more, changing the scope of study area for a city, the result will change accordingly. This gives rise to a dilemma of fractal dimension calculation, that is, fractal dimension values depend on both methods adopted and scope of study area defined in an empirical analysis. The main factors influencing fractal dimension calculation include prefractal structure, multifractal patterns, and self-affine fractal growth. The concrete reasons can be summarized as follows. **First, random prefractal structure result in diversity of fractal dimension estimation based on different methods and the deviation of estimated fractal dimension values from real fractal dimension values**. The regular monofractal dimension can be determined by its prefractal, but the dimension of a random fractal cannot be evaluated by its prefractal structure. The precondition of calculating its real or exact fractal dimension values are the process of linear scales of spatial measurement approach to infinitely small size. If the linear size of spatial measurement is small enough, different methods will lead to the similar or even the same fractal dimension values. Unfortunately, due to the limited scaling ranges of random prefractals, the linear size of spatial measurement is confined to certain range. **Second, random multifractal patterns results in the deviation of estimated fractal dimension values from the real fractal dimension values and the dependence of fractal dimension values on the scope of study area**. On the one hand, the global fractal dimension of a (regular or random) multifractal system cannot be determined by its prefractal structure. But in practice, we can only face the prefractal structure rather than real fractal structure of random multi-scaling fractals. On the other, multifractals bear spatial heterogeneity and different parts have different local fractal dimension values. Consequently, changing the size or the central location of study area results in different fractal dimension calculation results. **Third, self-affine fractal growth influences the fractal dimension estimation.** A self-similar growing fractal bears isotropic pattern and its fractal dimension can be estimated by area-radius scaling or density-radius scaling. However, many fractal growing processes of cities take on self-affinity and anisotropy. In this case, the scaling range often break into two segments, and it is difficult to find the reliable fractal dimension values. Because of interaction between random patterns, prefractal structure, multi-scaling processes, and self-affine growth, things become very complicated and the fractal dimension values take on diversity in an urban study. The dependence of fractal dimension values on scope of study area is indeed a problem, but strictly speaking, the dependence of fractal dimension on methods is not a problem. In mathematical modeling and quantitative analysis, the method-dependence of model parameter values is a common phenomenon. The solution to the problems lie in two aspects. On the one hand, find the most proper method for the special aspect of a city fractal and for the special direction of a study; on the other, use the comparable fractal dimension values to replace the real or exact fractal dimension values.

## Acknowledgements

This research was sponsored by the National Natural Science Foundations of China (Grant No. 41590843 & 41671167). The supports are gratefully acknowledged.